\documentclass[twocolumn,showpacs,preprintnumbers,amsmath,amssymb,superscriptaddress]{revtex4}

\usepackage{graphicx}
\usepackage{dcolumn}
\usepackage{bm}

   \newcommand{\euler}[1]{{\usefont{U}{eur}{m}{n}#1}}
    
    \newcommand{\umu}{\mbox{\euler{\char22}}}

\begin{document}


\title{$\rho^{0}$ Photoproduction in Ultra-Peripheral Relativistic Heavy Ion Collisions with STAR}

\affiliation{}
\affiliation{Argonne National Laboratory, Argonne, Illinois 60439}
\affiliation{University of Birmingham, Birmingham, United Kingdom}
\affiliation{Brookhaven National Laboratory, Upton, New York 11973}
\affiliation{California Institute of Technology, Pasadena, California 91125}
\affiliation{University of California, Berkeley, California 94720}
\affiliation{University of California, Davis, California 95616}
\affiliation{University of California, Los Angeles, California 90095}
\affiliation{Universidade Estadual de Campinas, Sao Paulo, Brazil}
\affiliation{Carnegie Mellon University, Pittsburgh, Pennsylvania 15213}
\affiliation{University of Illinois at Chicago, Chicago, Illinois 60607}
\affiliation{Creighton University, Omaha, Nebraska 68178}
\affiliation{Nuclear Physics Institute AS CR, 250 68 \v{R}e\v{z}/Prague, Czech Republic}
\affiliation{Laboratory for High Energy (JINR), Dubna, Russia}
\affiliation{Particle Physics Laboratory (JINR), Dubna, Russia}
\affiliation{University of Frankfurt, Frankfurt, Germany}
\affiliation{Institute of Physics, Bhubaneswar 751005, India}
\affiliation{Indian Institute of Technology, Mumbai, India}
\affiliation{Indiana University, Bloomington, Indiana 47408}
\affiliation{Institut de Recherches Subatomiques, Strasbourg, France}
\affiliation{University of Jammu, Jammu 180001, India}
\affiliation{Kent State University, Kent, Ohio 44242}
\affiliation{University of Kentucky, Lexington, Kentucky, 40506-0055}
\affiliation{Institute of Modern Physics, Lanzhou, China}
\affiliation{Lawrence Berkeley National Laboratory, Berkeley, California 94720}
\affiliation{Massachusetts Institute of Technology, Cambridge, MA 02139-4307}
\affiliation{Max-Planck-Institut f\"ur Physik, Munich, Germany}
\affiliation{Michigan State University, East Lansing, Michigan 48824}
\affiliation{Moscow Engineering Physics Institute, Moscow Russia}
\affiliation{City College of New York, New York City, New York 10031}
\affiliation{NIKHEF and Utrecht University, Amsterdam, The Netherlands}
\affiliation{Ohio State University, Columbus, Ohio 43210}
\affiliation{Panjab University, Chandigarh 160014, India}
\affiliation{Pennsylvania State University, University Park, Pennsylvania 16802}
\affiliation{Institute of High Energy Physics, Protvino, Russia}
\affiliation{Purdue University, West Lafayette, Indiana 47907}
\affiliation{Pusan National University, Pusan, Republic of Korea}
\affiliation{University of Rajasthan, Jaipur 302004, India}
\affiliation{Rice University, Houston, Texas 77251}
\affiliation{Universidade de Sao Paulo, Sao Paulo, Brazil}
\affiliation{University of Science \& Technology of China, Hefei 230026, China}
\affiliation{Shanghai Institute of Applied Physics, Shanghai 201800, China}
\affiliation{SUBATECH, Nantes, France}
\affiliation{Texas A\&M University, College Station, Texas 77843}
\affiliation{University of Texas, Austin, Texas 78712}
\affiliation{Tsinghua University, Beijing 100084, China}
\affiliation{Valparaiso University, Valparaiso, Indiana 46383}
\affiliation{Variable Energy Cyclotron Centre, Kolkata 700064, India}
\affiliation{Warsaw University of Technology, Warsaw, Poland}
\affiliation{University of Washington, Seattle, Washington 98195}
\affiliation{Wayne State University, Detroit, Michigan 48201}
\affiliation{Institute of Particle Physics, CCNU (HZNU), Wuhan 430079, China}
\affiliation{Yale University, New Haven, Connecticut 06520}
\affiliation{University of Zagreb, Zagreb, HR-10002, Croatia}

\author{B.I.~Abelev}\affiliation{University of Illinois at Chicago, Chicago, Illinois 60607}
\author{M.M.~Aggarwal}\affiliation{Panjab University, Chandigarh 160014, India}
\author{Z.~Ahammed}\affiliation{Variable Energy Cyclotron Centre, Kolkata 700064, India}
\author{B.D.~Anderson}\affiliation{Kent State University, Kent, Ohio 44242}
\author{D.~Arkhipkin}\affiliation{Particle Physics Laboratory (JINR), Dubna, Russia}
\author{G.S.~Averichev}\affiliation{Laboratory for High Energy (JINR), Dubna, Russia}
\author{Y.~Bai}\affiliation{NIKHEF and Utrecht University, Amsterdam, The Netherlands}
\author{J.~Balewski}\affiliation{Indiana University, Bloomington, Indiana 47408}
\author{O.~Barannikova}\affiliation{University of Illinois at Chicago, Chicago, Illinois 60607}
\author{L.S.~Barnby}\affiliation{University of Birmingham, Birmingham, United Kingdom}
% ...

\collaboration{STAR Collaboration}\noaffiliation


\date{\today}

\begin{abstract}

Photoproduction reactions  occur when  the electromagnetic field  of a
relativistic  heavy ion interacts  with another  heavy ion.   The STAR
collaboration  presents   a  measurement  of   $\rho^{0}$  and  direct
$\pi^{+}\pi^{-}$  photoproduction  in  ultra-peripheral  relativistic
heavy  ion  collisions at  $\sqrt{s_{NN}}$=200~GeV.   We observe  both
exclusive  photoproduction and  photoproduction accompanied  by mutual
Coulomb   excitation.   We    find   a   coherent   cross-section   of
 $\sigma(\text{AuAu} \rightarrow \text{Au}^{*}\text{Au}^{*}\ \rho^{0})$ =  530 $\pm$ 19 (stat.) $\pm$ 57 (syst.)~mb,  
 in accord with theoretical calculations
based on a Glauber approach, but considerably below the predictions of
a  color dipole  model.   The $\rho^{0}$  transverse  momentum
spectrum ( $p_{T}^{2}$ ) is  fit by a  double exponential curve including  both coherent
and   incoherent   coupling   to   the   target   nucleus;   we   find
$\sigma_{\text{inc}}/\sigma_{\text{coh}}  =  0.29  \pm  0.03$  (stat.)
$\pm$  0.08   (syst.).   The  ratio  of   direct  $\pi^{+}\pi^{-}$  to
$\rho^{0}$  production is comparable  to that  observed in  $\gamma p$
collisions at  HERA, and appears  to be independent of  photon energy.
 Finally, the measured $\rho^{0}$ spin helicity matrix elements agree within 
errors with the expected s-channel helicity conservation.

\end{abstract}

\maketitle

\section{\label{sec1}Introduction}

In heavy ion collisions when  the electromagnetic field of one nucleus
interacts   with  another  nucleus,
photoproduction can occur~\cite{baur, bertkn}.  Photoproduction is visible 
in  ultra-peripheral collisions  (UPCs), which  occur when  the impact
parameter  $b$ is  more than  twice the  nuclear radius  $R_A$,  so no
hadronic  interactions  are present.  The electromagnetic
 field of a relativistic nucleus may  be
represented as  a flux of  almost-real virtual photons,  following the
Weizs\"{a}cker-Williams method~\cite{ww}. In this framework,
the  physics of  the interactions  between  particles is equivalent to  that of  the
interactions between photons and  particles. The photon flux scales as
the  square of the  nuclear charge  and so  the cross-sections  can be
large in heavy ion interactions.

Photoproduction  of $\rho^0$s  occurs when  a photon from one nucleus fluctuates  to a
quark-antiquark pair, which then  scatters elastically from the other
nucleus,  emerging as  a $\rho^{0}$.   The elastic  scattering  can be
treated as being due to Pomeron exchange~\cite{forross}.

The  cross-section  for  $\rho^{0}$   production  depends  on  the  $q
\overline  q$   coupling  to  the  nuclear   target.   For  $\rho^{0}$
production  at large  transverse momentum,  $p_T$, the  $q\overline q$
pair couples to the individual nucleons.
  The incoherent cross-section scales roughly as
the atomic  number $A$, minus a correction due to  nuclear absorption of
the $\rho^{0}$.

At smaller $p_T$,  roughly $p_T < \hbar/R_A$, the  $q\overline q$ pair
couples coherently  to the  entire nucleus; naively,  this leads  to a
cross-section that  scales as $A^2$.   
 The coherent production  is regulated by  the nuclear form  factor $F(t)$,
so   $\rho^{0}$    photoproduction   is   sensitive    to   the
$\rho^{0}$-nucleon interaction cross-section and the nuclear structure
functions~\cite{bertkn}.

There  are  three  published  calculations of  the  coherent  $\rho^0$
photoproduction  cross-section  in heavy  ion  collisions.  The  first
model (Klein and Nystrand - KN), uses vector meson dominance (VMD) plus a
classical mechanical Glauber approach  for nuclear scattering. KN uses
information  from  the  $\gamma  p  \rightarrow  Vp$  experiments  for
extrapolation~\cite{ksjn}.  The model predicts a  total  coherent  $\rho^{0}$
photoproduction cross-section  $\sigma_{\rho^{0}}$=590~mb in   gold-gold    collisions   at
$\sqrt{s_{NN}}$=200~GeV.  The second
model  (Frankfurt,  Strikman and  Zhalov  - FSZ)  treats  the  $\rho^{0}$
production   using   the  generalized   quantum   VMD   and  the   QCD
Gribov-Glauber    approach~\cite{fsz,    fsz200};    it    predicts
$\sigma_{\rho^{0}}$=934 mb~\cite{fsz200},
about 60~\%  higher than  the KN  model, but with  a similar
rapidity  distribution.  The  third  model (Goncalves and  Machado -  GM)
describes the photoproduction of the vector mesons in UPC events using
the QCD  color dipole approach~\cite{gm}. This  model includes nuclear
effects    and     parton    saturation    phenomena.      It    finds
$\sigma_{\rho^{0}}$=876~mb, with a rapidity distribution very different from that 
of the other models.  The FSZ  and GM models  provide predictions for the  momentum transfer
dependence   of   both   coherent  and   incoherent   $\rho^{0}$
production.

The $\rho^{0}$ photoproduction on  nuclear targets has been studied at
fixed-target experiments~\cite{fixedtarget} and at RHIC (Relativistic
Heavy Ion Collider).
Previous fixed-target photoproduction experiments with nuclear targets
were done at relatively low collision energies~\cite{fixedtarget}.
The PHENIX collaboration has studied $J/\psi$ photoproduction~\cite{phenix} in heavy ion interactions at RHIC. 
The  Solenoidal  Tracker  at RHIC (STAR)  collaboration has  published
measurements   of   the   $\rho^0$~production   cross-section   at   a
center-of-mass energy of  $\sqrt{s_{N N}} = 130$~GeV per nucleon pair~\cite{meis}. This
work presents results at a higher center-of-mass energy of $\sqrt{s_{N
    N}}  = 200$~GeV.  To  produce a  meson  with mass~$m_{\rho^{0}}$  at
rapidity~$y_{\rho^{0}}$  a minimum photon-nucleon  center-of-mass energy
of  $W_{\gamma N} =  [ \sqrt{s_{N  N}} \,  m_{\rho^{0}} \exp(y_{\rho^{0}})
]^{1/2}$  is  required.  At  mid-rapidity this  corresponds  to  about
12.5~GeV, and  in the region  $|y_{\rho^{0}}| < 1$  to a range of  $7.6 <
W_{\gamma  N}  <  20.6$~GeV,  which  is well  above  that  of  previous
fixed-target     photoproduction      experiments~\cite{fixedtarget}.

The $\gamma$-nucleon  collision energy is related to  a minimum photon
energy in the lab frame which is given by

\begin{equation}\label{photonenergyrap}
  E_\gamma = \frac{m_{\rho^{0}}}{2} \exp(\pm y_{\rho^{0}})
\end{equation}

The two  signs reflect  the ambiguity about which nucleus emits the photon.

In the rest frame of the target nucleus, the minimum photon energy  is  $2\gamma_L$~times 
higher than in the lab frame, where $\gamma_L$ is the Lorentz boost of the beam.  For $\sqrt{s_{N N}}$ = 200~GeV, 
 $\gamma_{L}$ is about 108; at mid-rapidity, this corresponds to a photon energy in the target frame of about 84~GeV.  
For non-zero rapidities, most of the $\rho^{0}$ production comes from the solution with the lowest energy, so the minimum 
energy required is even less.  To evaluate whether or not the $\rho^{0}$s can be produced coherently, one must compare this energy 
to the maximum photon energy.  In the rest frame of the target nucleus, this maximum is determined by 
the uncertainty relation $E_{\gamma}^{target} \lesssim  (2 \gamma_L^2 -
1) \, \hbar / R_A$.  For $\gamma_L$ = 108, this maximum is approximately 650~GeV.  Since the minimal energy 
required to produce a $\rho^{0}$ is much less than 650~GeV, we expect to easily observe coherently 
produced $\rho^{0}$ mesons.

\begin{figure}[htb]
\begin{picture}(100,100)
\put(-130,-400){\includegraphics{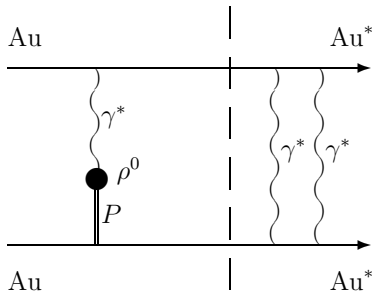}}
\end{picture}
\caption{\label{fig:diagram}     The     diagram    for     $\rho^{0}$
  photoproduction  accompanied  by  mutual  Coulomb  excitation.   The
  latter  process proceeds  by  mutual photon  exchange; the  vertical
  dashed line shows how the photoproduction factorizes from the mutual
  Coulomb excitation.}
\end{figure}

In this study, we have collected data at higher energy ($\sqrt{s_{NN}}
=  200$~GeV vs. $\sqrt{s_{NN}}  = 130$  GeV) and  with about  10 times
more statistics  than  the previous  STAR  study,  allowing  for more  precise
measurements  of  the cross-section.  We  have  extended the  previous
 analysis  by   measuring  both   the   coherent  and   incoherent
contributions to  the photoproduction  cross-section and we  have also
measured the spin-matrix elements for $\rho^{0}$ production.
 In addition  to exclusive $\rho^{0}$ photoproduction,  we have studied
$\rho^{0}$ photoproduction  accompanied by mutual  Coulomb excitation,
as is shown in  Fig. \ref{fig:diagram}.  This process primarily occurs
via 3-photon  exchange, with one photon producing  the $\rho^{0}$, and
one  photon   exciting  each  nucleus~\cite{nbk,multiphoton}.   Each
single-photon reaction  is independent, and the  cross-sections may be
written as an integral over the impact parameter

\begin{eqnarray}\label{eq:crossprob}
\lefteqn{\sigma(\text{Au Au} \rightarrow \text{Au}^{*}\text{Au}^{*} \rho^{0})= } \nonumber\\ & & {} = \int d^{2}b  
\big[1- P_{\text{Had}}(b)\big] P_{\rho^{0}}(b)P_{Xn,1}(b)P_{Xn,2}(b)
\end{eqnarray}
where   $P_{\text{Had}}(b)$   is  the   probability   of  a   hadronic
interaction,   $P_{\rho^{0}}(b)$  is  the   probability  to   produce  a
$\rho^{0}$, and $P_{Xn,1}(b)$  and $P_{Xn,2}(b)$ are the probabilities
to  excite nucleus 1  and 2  respectively.  The  three-photon exchange
reactions   are   biased  toward   smaller   impact  parameters   than
 single-photon reactions,  leading to a  harder photon spectrum  and an
altered  rapidity  distribution.  In  mutual Coulomb  excitation,  the
nuclei decay  primarily by channels involving  neutron emission.  This
is  attractive  experimentally,  since  the  neutrons  provide  simple
trigger  signals  that can  be  detected  with  the STAR Zero  Degree
Calorimeters (ZDCs)~\cite{zdcdes}.

One   particular   nuclear   excitation   merits   special   interest:
electromagnetic    excitation   to    a    Giant   Dipole    Resonance
(GDR)~\cite{baurbertul}  of either  one  or both  ions; GDR  involves
particularly low-energy photons.
 A single GDR is the main contribution
in the total fragmentation cross-section induced by Coulomb excitation in  UPCs.
 GDRs  usually  decay  by single  neutron  emission, which  is
considered to  be a  major source of beam loss in  heavy ion
colliders~\cite{bertkn}.

The  hypothesis of $s$-channel  helicity conservation  for vector
mesons  suggests  that  the  vector meson produced in the collision  will  retain  the
helicity  of   the  initial  photon~\cite{gilman}.    The  differential
production  cross-section and  the decay  angular distribution  of the
vector meson can  be expressed as a function of  the vector meson spin
density matrix, which is represented by the sum of the helicity states
and  encompasses  transverse   and  longitudinal  elements  and  their
combinations~\cite{schil}. As a  consequence of parity conservation
and the symmetry properties  of the  matrix elements,  the  decay angular
distribution can be greatly simplified. If the helicities
are conserved  in the hadronic  center-of-mass system, there  are only
three   independent  helicity   amplitudes  in   the   final  function
~\cite{schil}. In  this paper, we  report a measurement of  these three
$\rho^{0}$ spin density  matrix elements for $p_{T} <  $ 150~MeV/c and
a  photon-nucleon   center-of-mass  system  energy   $W_{\gamma  N}$
$\approx$  10~GeV,  beyond   previous  fixed-target
experiments~\cite{fixedschc}.

\section{\label{sec2}Experimental Setup and Triggering}

This analysis  uses data  taken with the STAR detector at Brookhaven  National Laboratory  during the  2001 run.
Gold  nuclei were collided  at  $\sqrt{s_{NN}}$ = 200~GeV  and  the  charged
particle tracks  were reconstructed  in a cylindrical  Time Projection
Chamber (TPC)~\cite{tpcdes}.  The TPC is  a 4.2~m long  barrel with a
2~m radius operated  in a solenoidal magnetic field  of 0.5~T. The TPC
detected charged  tracks with pseudorapidity $|\eta| < 1.2$  and $p_T >
 100$~MeV/c with  an efficiency of  about 85~$\%$.  The TPC is  surrounded by
240 Central Trigger Barrel (CTB)~\cite{trigdes} slats which are plastic
scintillator detectors spaced every  6 
 degrees in $\Phi$ with complete (hermetic) coverage  over the  full range 
 of  rapidity covered  by the TPC.   Two Zero  Degree Calorimeters  
 (ZDCs)~\cite{zdcdes}  are situated along the  beam pipe  at $\pm$ 18~m 
from the interaction  point. They have an  acceptance close to  unity for 
  the neutrons  originating from nuclear break-up.

This analysis  used data from two  triggers: a topology  trigger and a
minimum bias trigger.  The topology trigger divided the CTB detector into four azimuthal quadrants.
  A coincidence between the
left and right side quadrants was required, while at the same time the
top and bottom quadrants were required to be empty.  The veto from the
top and  bottom quadrants was used  to reduce the trigger  rate due to
cosmic rays.

\begin{figure}[htb]
\begin{picture}(100,100)
\put(-80,-10){\includegraphics{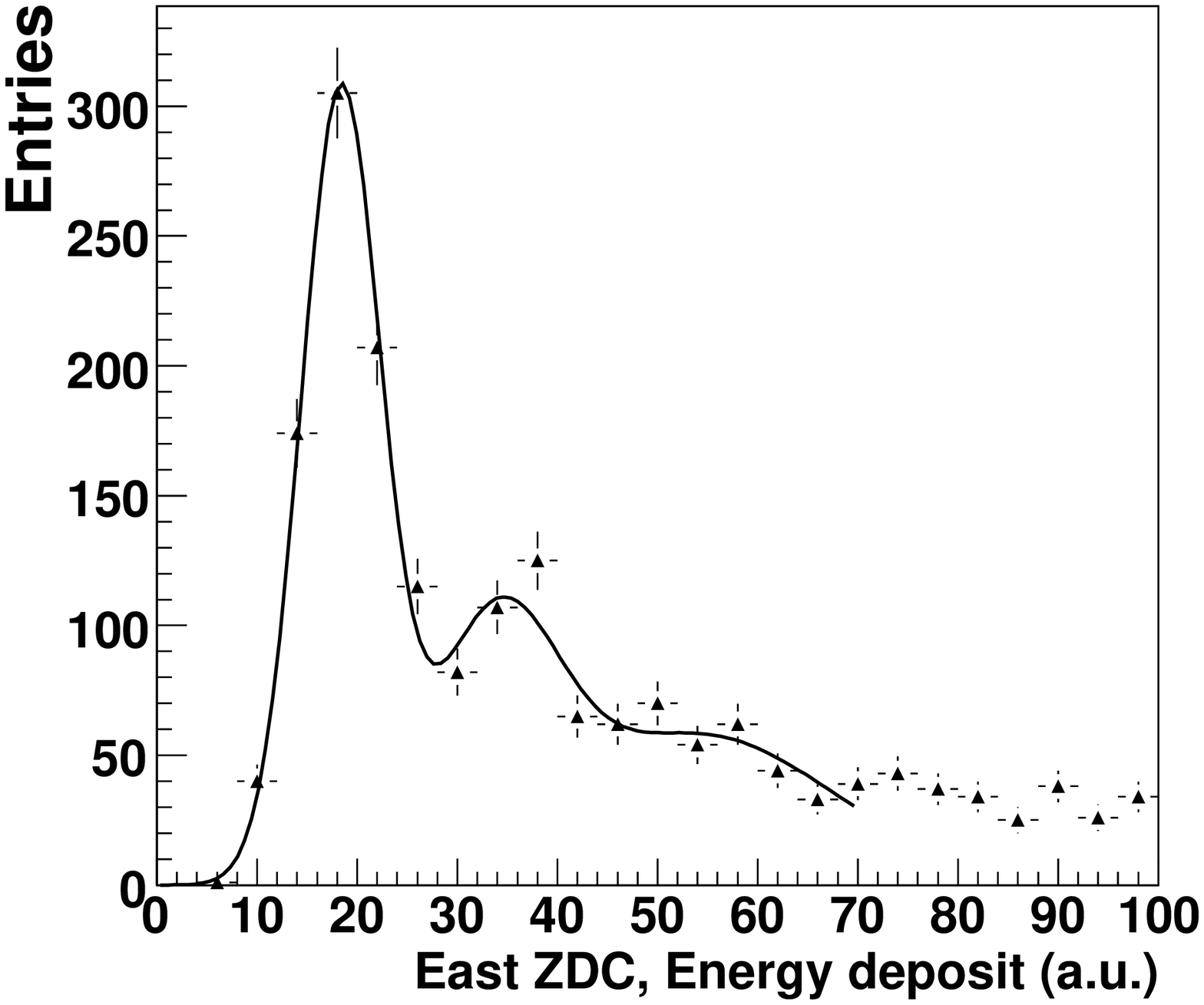}}
\put(50,-10){\includegraphics{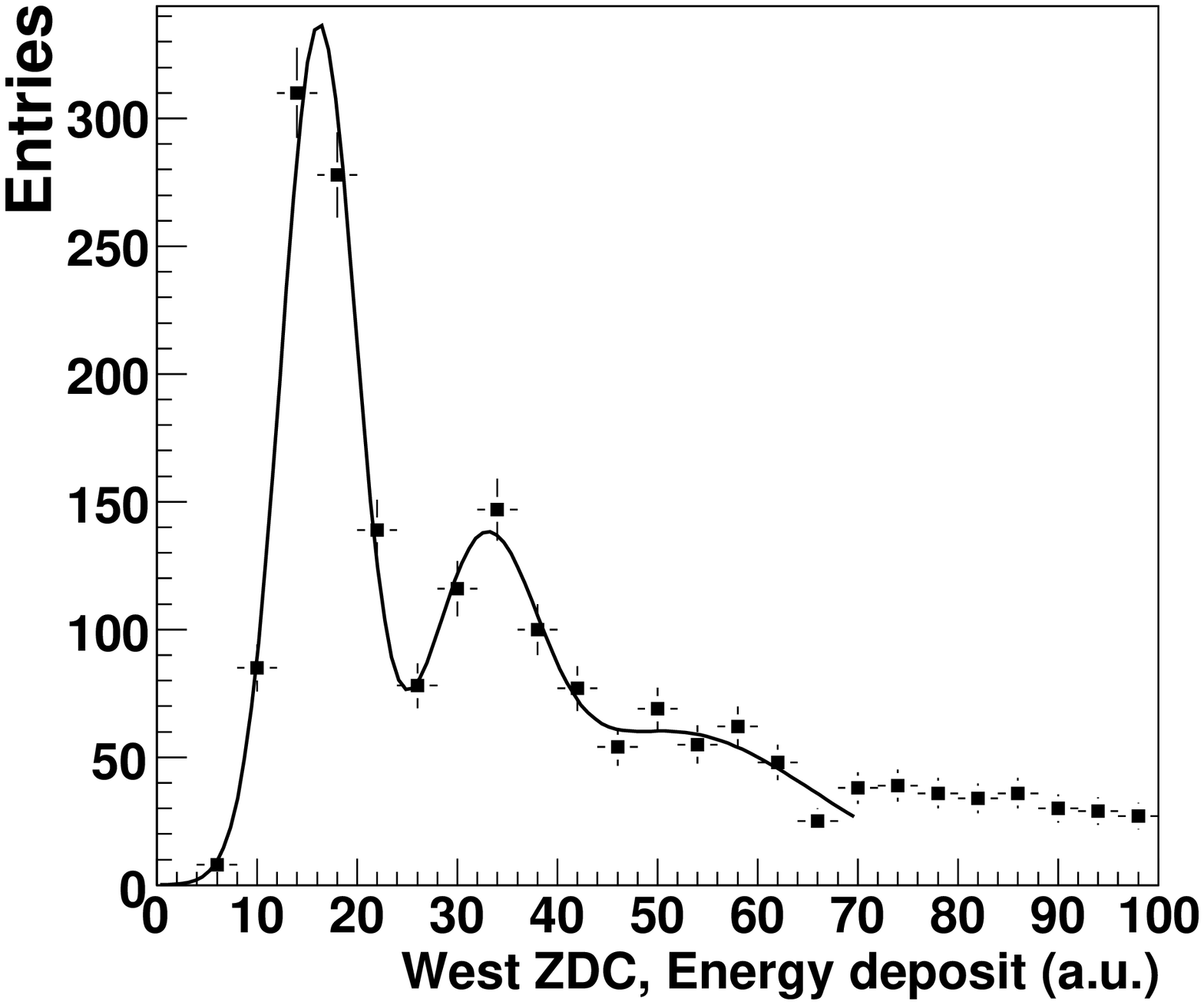}}
\put(-45,10){1n}
\put(-27,10){2n}
\put(-5,10){3n}

\put(83,10){1n}
\put(100,10){2n}
\put(120,10){3n}

\end{picture}
\caption{\label{fig:zdc}  ZDC spectra obtained  with the  minimum bias
  sample after the $\rho^{0}$ selection cuts are applied, and fit with
  three Gaussians. The east ZDC is  shown on the left and the west ZDC
  is shown  on the  right. The  ratio of numbers  of candidates  in the
  West ZDC of 1n:2n:3n is 1:~0.48~$\pm$~0.03:~0.42~$\pm$~0.03, while in the East ZDC, we find 1n:2n:3n is
1:~0.46~$\pm$~0.03:~0.42~$\pm$~0.03.}
\end{figure}

The minimum bias  trigger required a coincidence between  the two ZDCs
and  thus  was  sensitive  to photoproduction  accompanied  by  mutual
Coulomb excitation.   By eliminating cosmic rays  and other extraneous
interactions,  this trigger had  considerably better  selectivity than
the topology  trigger.  The ZDCs have sufficient  energy resolution to
count the number of neutrons  emitted by the outgoing gold nuclei.  We
distinguish between  several different  excitation modes: $XnXn$  - at
least one neutron  in each of the ZDC detectors,  $1n1n$ - exactly one
neutron in each of the ZDC detectors, $0nXn$ - at least one neutron in
one  of the  ZDC detectors  and none  in the  other, and  $0n0n$  - no
neutrons in either  ZDC.  The last two modes  are only accessible with
the   topology  trigger.   A  typical   ZDC  spectrum   is   shown  in
Fig.   \ref{fig:zdc}. In the West ZDC, the
ratio of 1n:2n:3n is 1:~0.48~$\pm$~0.07:~0.42~$\pm$~0.04, while in the East ZDC, we find 1n:2n:3n is
1:~0.46~$\pm$~0.08:~0.42~$\pm$~0.04. This   spectrum  allows   us  to   measure  the
cross-section for different excitation states.

\section{\label{sec3}$\rho^{0}$ photoproduction}

\subsection{\label{sub:sec2}Event Selection}

This  analysis  selected events  with  two  oppositely charged  tracks
forming a primary vertex  (beam interaction point) 
and having  less than six  reconstructed charged tracks per  event.  A
$\rho^{0}$ photoproduction event should have exactly two tracks in the
TPC,  but additional  tracks may  come from  overlapping interactions,
including beam-gas events.  The STAR  TPC has a 36~$\umu$s drift time,
so any charged particles traversing  the TPC within $\pm 36~\umu$s may
deposit  energy  which  overlaps  with  the tracks  of  interest.   We
accounted for the  effect of these tracks in  our analysis by allowing
for varying numbers of total tracks in the event, including 
primary and secondary tracks. 
We analyze only events with exactly 2 tracks forming a primary vertex. 
 If  the cut  on the total  number of  tracks is  relaxed  from 2  to 5,  the number of reconstructed $\rho^{0}$ 
 increases by 27 $\pm$ 1~\%. The results were corrected
by this factor.

The reconstruction software formed a vertex from the
charged particle
trajectories.   An iterative procedure was used to successively remove
tracks that were inconsistent with
the vertex position; after the least consistent track was removed, the a
new vertex was found.  This process
continued until a vertex was found with an acceptable probability.  
 The  single  track  reconstruction  efficiency for  $|\eta|<1.2$  is  about
85~$\%$, and the  vertex finding efficiency for a  two-track vertex is
80 $\pm$  2~$\%$. There are  several types of  backgrounds: peripheral
hadronic interactions, other photonuclear interactions, $e^+e^-$ pairs
from two-photon interactions, and unrelated processes such as beam-gas
interactions, cosmic ray muons and pile-up events.
 These backgrounds  can be reduced  by cuts on the  total multiplicity,
vertex position, and other event characteristics.

The  multiplicity cut  suppresses the  contribution from  hadronic and
pile-up events. After the cuts on the multiplicty, the minimum bias
 and topology samples contain 48,670 and 98,112 events, respectively. 
In order to  reduce the backgrounds  originating from
processes like beam-gas events, upstream interactions, cosmic rays and
pile-up events,
 we  selected events with  primary vertices  within 15~cm  radially and
100~cm longitudinally (along the beam  direction) of the center of the
interaction region. Those two cuts reject approximately 25~$\%$ of the
events.
 We also required that tracks have at  least 14 hits in the TPC (out of
a maximum of 45 possible hits). This cut rejected another 30~$\%$ of the events.
In order  to retain  as many of  the incoherently  produced $\rho^{0}$
mesons  as   possible  while  removing   combinatorial  background,  a
relatively soft cut  on the $\rho^{0}$ transverse momentum ($p_T  \le$ 550 MeV/c)
was applied. 
 After cuts, the minimum bias
sample contains 5,011 events, while the 
topology sample contains 14,693 events.     

Backgrounds from two-photon interactions and non-$\rho^{0}$ photonuclear
interactions are small.   The cross-sections for two-photon production
of $e^+e^-$ in the STAR acceptance are small~\cite{epem}.  They were a
small  correction in  the  130  GeV analysis,  but  the current  study
requires  $M_{\pi^{+}\pi^{-}}$ $>$  500  MeV/c$^2$.  With  this  cut, the corrections are
negligible. 

 There is  also  a small
background  from coherent  $\omega$ photoproduction;  the $\pi^{+}\pi^{-}$
final state has a  negligible 2.2~$\%$ branching ratio~\cite{pdg}, but
the $\pi^{+}\pi^{-}\pi^0$ final state  has an 89~$\%$ branching ratio. 
The ($\pi^{+}\pi^{-}$) invariant mass distribution from 3-pion decay peaks at a lower value, and higher $p_T$  
than the $\rho^{0} \rightarrow\pi^{+}\pi^{-}$.  This is about  a 2.7~$\%$  correction to
 the incoherent $\rho^{0}$ cross-section; we neglect this here.

 The hadronic  interactions produce  much higher  multiplicity final  states than the photoproduced $\rho^{0}$ and can be easily 
 distinguished  by their total multiplicity~\cite{kn}.

Even  with the  veto from  the  top and  bottom CTB  quadrants in  the
trigger, some cosmic-rays remain in  the topology sample. Particles that pass
near  the  interaction  region  may  be reconstructed  as  a  pair  of
back-to-back  tracks  with  net  charge  0,  net  $p_T\approx  0$  and
$y_{\rho^{0}}\approx  0$.   These events  are  removed  by
applying  a cut  on  the rapidity  so  that the  accepted events  have
$|y_{\rho^{0}}|>0.01$.
On  the other hand,  the ZDC  energy coincidence  requirements largely
eliminate cosmic-ray contamination in the minimum bias sample.

We use two approaches to  estimate the remaining backgrounds.  
As with the 130~GeV analysis,  like-sign pairs ($\pi^+\pi^+$ and $\pi^-\pi^-$)
provide  a  good background  model~\cite{meis}.   That analysis  only
considered coherent $\rho^{0}$  production; the like-sign background was
scaled up  by a factor  of 2.2  to match the  data at high  $p_T$.  By
definition,  this   treats  incoherent  $\rho^{0}$   production  as  a
background.  We  use this approach to measure  the ratio of
directly  produced $\pi^{+}\pi^{-}$ pairs  to $\rho^{0}$  production ($|B/A|$ ratio)
  for the  coherently  produced $\rho^{0}$  mesons, since  it
correctly estimates the combinatorial background.

For the rest of the
measurements, we  use the unscaled  background in order to  retain the
incoherent $\rho^{0}$ signal. For incoherent
$\rho^{0}$ photoproduction, we split the invariant
mass histogram into different $p_T$ ranges, and fit each $p_T$ bin
separately to determine the yield.  
In  our  fits  to  the  $M_{\pi^{+}\pi^{-}}$ spectrum  the  background  is
parameterized by a polynomial.
The polynomial function is fixed with parameters obtained from the fit
of   the    polynomial   function   to    the   non-scaled   like-sign
distribution.   These   different   approaches  for   the   background
description
cause a 3~$\%$ systematic error in the cross-section measurement.

\subsection{\label{sub1:sec3}Efficiency and Acceptance Determination}

The acceptance of the detector  was studied using a Monte Carlo event generator which is based on
the KN model~\cite{nbk, ksjn} to generate events
 which  reproduce the  kinematic  properties and spatial distributions of  the
$\rho^{0}$ mesons produced  via coherent photoproduction. These events
were passed  through a realistic detector  simulation which reproduces
detector  resolution  and  efficiency.   The efficiency  includes  the
detector  acceptance, track and  vertex reconstruction efficiencies, and
selection cuts.

The $\rho^{0}$ reconstruction efficiency  was studied as a function of 
$p_{T}$,  $p_{T}^{2}$,  $\Phi$  (azimuthal  angle  in  the  center-of-mass system of AuAu),
$\Theta$ (polar angle in the center-of-mass system of AuAu), $y_{\rho^{0}}$ (rapidity) and
$M_{\pi^{+}\pi^{-}}$.  The mean efficiency  for minimum bias $\rho^{0}$s with
$|y_{\rho^{0}}|<1$ is  44 $\pm$ 2~$\%$. This  efficiency is relatively
constant with  respect to  $p_{T}$ and azimuthal  angle, but  drops as
$|y_{\rho^{0}}|$ increases, due to the TPC acceptance dropping at higher
rapidity.   The mean efficiency  for topology-triggered  $\rho^{0}$s with
$|y_{\rho^{0}}|<1$ is 11 $\pm$ 1~$\% $,
 the efficiency drops slowly  as $p_T$ or $|y_{\rho^{0}}|$ rises. There
is  also an  azimuthal dependence  due  to the  topology trigger veto
regions.

The    estimated   resolution    for    $p_{T}$,   $y_{\rho^{0}}$    and
$M_{\pi^{+}\pi^{-}}$  are  approximately  6~MeV/c,  0.01  and  6~MeV/c$^2$
respectively  for  track  pairs  that passed  through  the  $\rho^{0}$
selection cuts.

\subsection{\label{sub2:sec3}Luminosity}

The  luminosity for  the minimum  bias  data sample  is calculated  by
assuming that the main  contribution to the total cross-section arises
from hadronic  production, with a known  cross-section. The luminosity
was measured by  counting events with at least  14 primary tracks with
$p_{T} \geq$ 0.1~GeV/c  and $| y_{\rho^{0}} | \leq$  0.5. These events
correspond to  80~$\%$ of the total  hadronic production cross-section
of 7.2~b~\cite{baltz}. 
An extra correction  is required to remove the  effects of an unstable
dead time caused by the  SVT (Silicon Vertex Detector). The integrated
luminosity of  the minimum bias sample  is measured to be  $L$ = 461 
mb$^{-1}$ with  a systematic uncertainty  of 10~$\%$ which  is largely
due to the uncertainty of the gold-gold hadronic cross-section. 

\subsection{\label{sub3:sec3}Invariant Mass Fit Function}

\begin{figure}[htb]
\begin{picture}(100,350)
\put(-45,160){\includegraphics{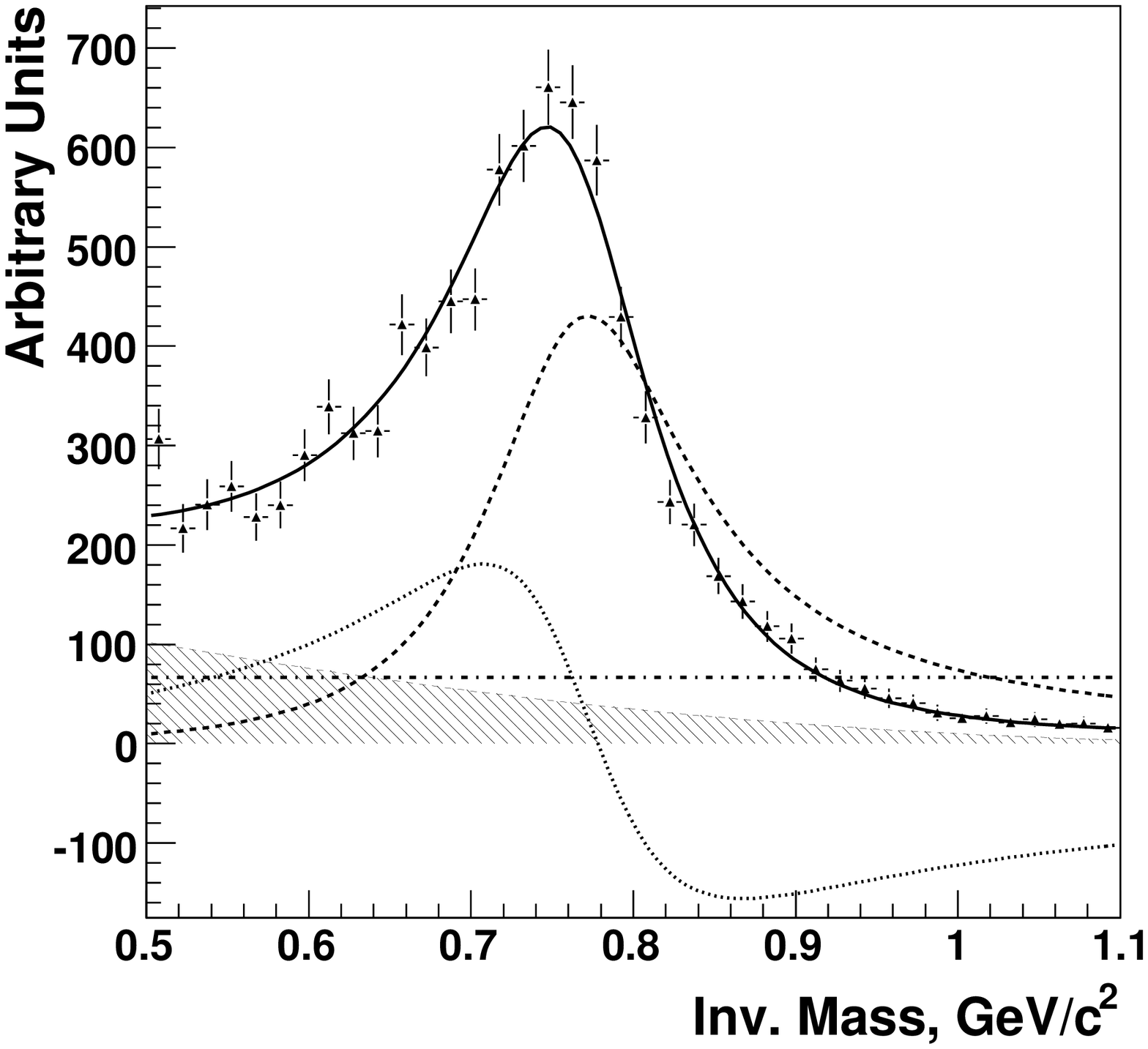}}
\put(-52,-10){\includegraphics{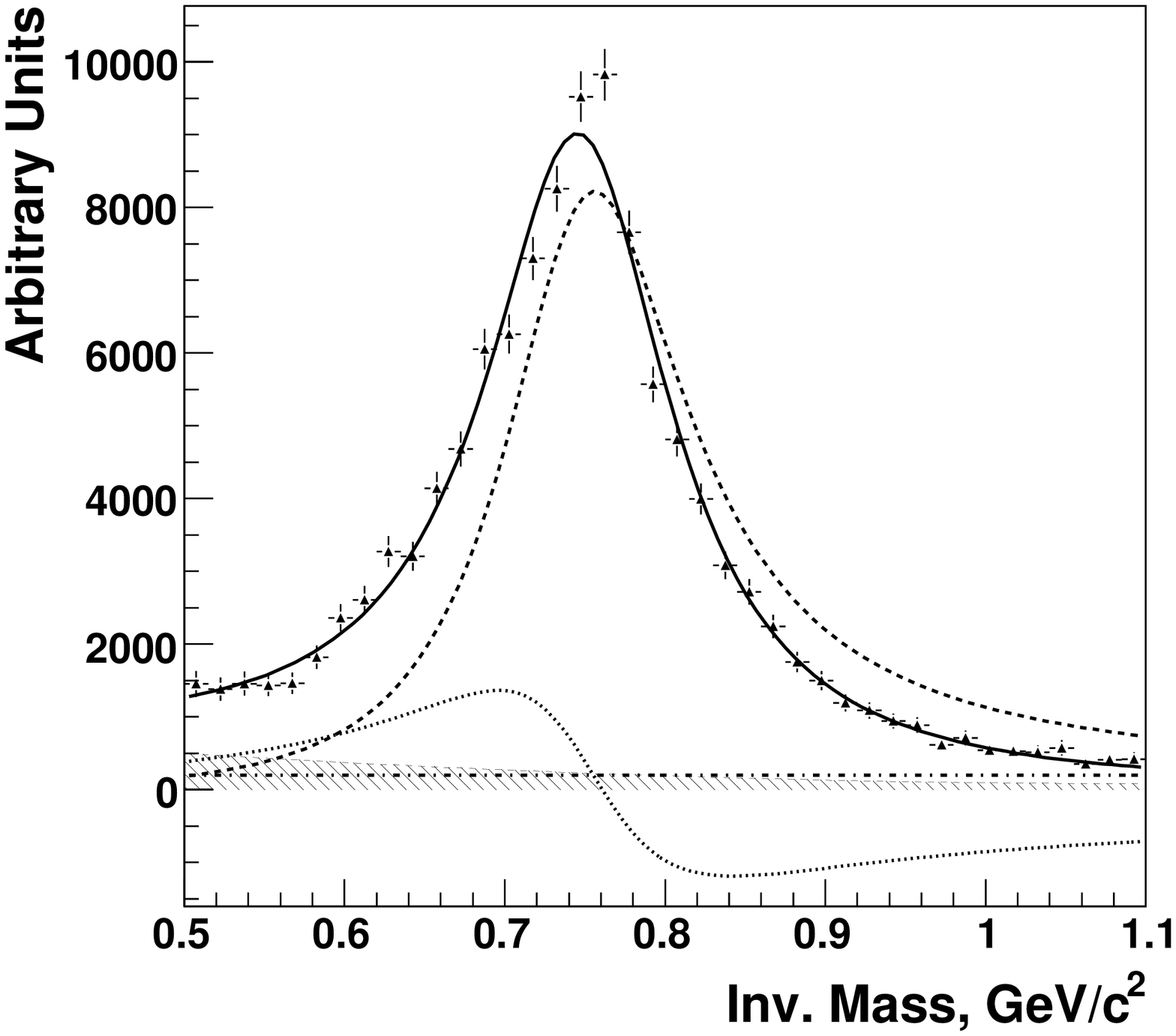}}
\end{picture}
\caption{\label{fig:mass}Top: The invariant  mass distribution for the
  coherently  produced  $\rho^{0}$ candidates  from  the minimum  bias
  sample  with  the cut  on the $\rho^{0}$ transverse momentum  $p_T  <$  150~MeV/c.   
  Bottom:   The  invariant  mass
  distribution  for  the  coherently  produced  $\rho^{0}$  candidates
  obtained  from  the topology  sample  with  the cut  on the $\rho^{0}$ transverse momentum $p_T  <$ 150~MeV/c.  The
  hatched  area  is   the  contribution  from  the  combinatorial
  background. The solid line corresponds to Eq. \ref{eq:fitfunc} which
  encompasses   the  Breit-Wigner   (dashed),  the   mass  independent
  contribution from the  direct $\pi^{+}\pi^{-}$ production (dash-dotted),
  and the interference term(dotted).}
\end{figure}

The invariant mass  distribution of track pairs was  found by assuming
that   all   reconstructed   particles   were   pions;   no   particle
identification  was  needed due  to  the  low  background level  after
selection cuts were applied to the tracks.
 The invariant mass distributions for the minimum bias and topology 
samples are shown in Fig. \ref{fig:mass}.      

Pion  pairs may  be  photoproduced via  an intermediate $\rho^{0}$, or  the
photon  may fluctuate  directly to  a $\pi^{+}\pi^{-}$ pair.  The  direct process
produces   a  flat   $M_{\pi^{+}\pi^{-}}$  mass   distribution.   The  two
experimentally   indistinguishable   processes   interfere   and   the
interference  is constructive  for $M_{\pi^{+}\pi^{-}}  <  M_{\rho^{0}}$ and
destructive for $M_{\pi^{+}\pi^{-}} > M_{\rho^{0}}$~\cite{ryskin}.

The invariant  mass distribution of the $\rho^{0}$  candidates was fit
with  a   relativistic  Breit-Wigner~\cite{breit}   function  plus  a
contribution  for  the   direct  $\pi^{+}\pi^{-}$  production  and  an
interference (S\"oding) term~\cite{soding, sakurai}.
 The background is described by  a 2nd order polynomial. An estimate of
the  background  from the  like-sign  pairs  was  used to  obtain  the
parameters for the polynomial function.

The fit function is:
\begin{equation}\label{eq:fitfunc}
\frac{dN}{dM_{\pi^{+}\pi^{-}}}=\left| A\frac{\sqrt{M_{\pi^{+}\pi^{-}}M_{\rho^{0}}\Gamma_{\rho^{0}}}}{M_{\pi^{+}\pi^{-}}^{2}-M_{\rho^{0}}^{2}+iM_{\rho^{0}}\Gamma_{\rho^{0}}}+B \right|^{2} + f_{p},
\end{equation}

where  \\ 

\begin{eqnarray}
\Gamma_{\rho^{0}}&=&\Gamma_{0}\cdot(M_{\rho^{0}}/M_{\pi^{+}\pi^{-}})
\nonumber \\
 &&\times\left[(M_{\pi^{+}\pi^{-}}^2-4m_{\pi}^2)/(M_{\rho^{0}}^2-4m_{\pi}^2)\right]^{3/2}  
\end{eqnarray}
is the momentum-dependent width  of the $\rho^{0}$ and $M_{\rho^{0}}$ is the
mass  of the  $\rho^{0}$,  $A$  is the  amplitude  for the  Breit-Wigner
function,  $B$  is  the  amplitude  for  the  direct  $\pi^{+}\pi^{-}$
production and $f_{p}$  is the fixed second order  polynomial which is
used  to  describe  the background.    For  the  minimum  bias  data  set,
$\Gamma_{\rho^{0}}$ = 0.162 $\pm$ 0.007~GeV/c$^2$ and $M_{\rho^{0}}$ = 0.775
$\pm$ 0.003~GeV/c$^2$ from the fit.  These values are in good agreement with the PDG~\cite{pdg}
values.  The difference between the yield obtained with fixed $\rho^{0}$
width and mass  position from that obtained without  fixing the width
and mass  is about 2~$\%$. Fixing the width  leads to an
increase in the $\chi^2/DOF$ of  5~$\%$. Using the above mentioned fit
procedure, the  minimum bias sample contains 3,075  $\pm$ 128 $\rho^{0}$
candidates,  while  the  topology  sample contains  13,054  $\pm$  124
$\rho^{0}$ candidates.

\begin{figure}[htb]
\begin{picture}(100,200)
\put(-60,0){\includegraphics{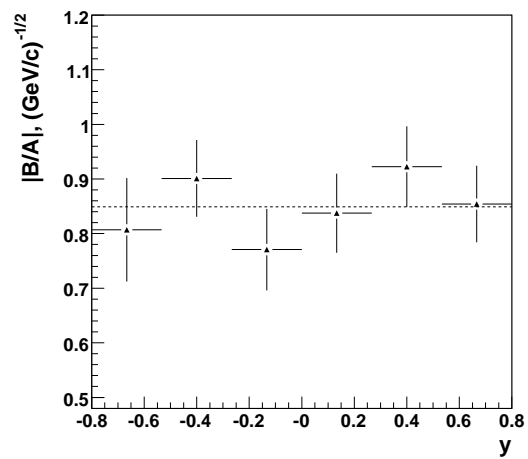}}
\end{picture}
\caption{\label{fig:barapidity} The ratio $|B/A|$ as a function of $y_{\rho^{0}}$ for the minimum bias data, 
obtained by fitting  Eq.\ref{eq:fitfunc} to 
the invariant mass distributions in bins of $y_{\rho^{0}}$.}
\end{figure}

For the minimum bias data, the measured value of $|B/A|$ is 0.89 $\pm$
0.08 (stat.) $\pm$ 0.09 (syst.)~(GeV/c)$^{-1/2}$;  the  systematic error  due to  the
background description is 3~$\%$.
 Figure   \ref{fig:barapidity}  shows  that   $|B/A|$  does   not  vary
significantly as a function of rapidity.  Since rapidity is related to
photon energy  (Eq. \ref{photonenergyrap}) this also  shows that there
is  no significant variation  with photon  energy.  The same energy-independence was seen by the
ZEUS experiment at HERA \cite{breit}, and is expected in Pomeron exchange. In contrast, at
lower energies \cite{fixedtarget}, the direct $\pi^+\pi^-$ component decreased as $W$ rose.  The
difference may be due to $\rho$ production by meson exchange;
the meson exchange component is expected to be negligible at RHIC energies. The observed $|B/A|$ ratio
is independent of the polar and azimuthal angle, as expected. The slight asymmetry in the distribution is 
believed due to differences between the two parts of the STAR TPC, and
is included within the systematic uncertainties.

Our measured value  for $|B/A|$ is in agreement  with the 130~GeV STAR
 result,  $|B/A|$   =  0.81  $\pm$   0.08  (stat.)  $\pm$  0.20 (syst.)~(GeV/c)$^{-1/2}$
~\cite{meis}. The ZEUS studies of $\gamma p\rightarrow \rho^{0} p$ find
$|B/A| = 0.67~\pm$ 0.03~(GeV/c)$^{-1/2}$
~\cite{breit},  but for a slightly different kinematic interval.  The ZEUS results are
for a momentum   transfer squared of  $t<0.5$~(GeV/c)$^2$. At mid-rapidity in a collider environment,
the longitudinal  momentum  transfer squared from  the  target  nucleus $t_{||}  =
(m_V^2/2E_{\gamma})^2$ is small ($\approx 2$~(MeV/c)$^2$).  
Therefore $t \approx t_\perp \approx p_T^2$ and so we can extrapolate
the ZEUS measurement of the $t$ dependence of $|B/A|$ down to our average value of
$t \approx p_T^2 < 0.015$~(GeV/c)$^2$.  Extrapolating the ZEUS results, we find $|B/A| \approx 0.8$, 
which is consistent with our results.
 The decrease of $|B/A|$ with increasing $|t|$ and the independence of the
polar and azimuthal angles is expected assuming that the form factor of
the vector meson depends only on $t$ and so 
no $\sqrt{s_{NN}}$ dependence is expected~\cite{ryskin}. 

In the rest of this paper, we will quote our results in terms of momentum transfer squared $t$.

\subsection{Cross-Sections $d\sigma/dy$ and $d^{2}\sigma/dydt$ for Minimum Bias events}

\begin{figure}[htb]
\begin{picture}(100,200)
\put(-60,0){\includegraphics{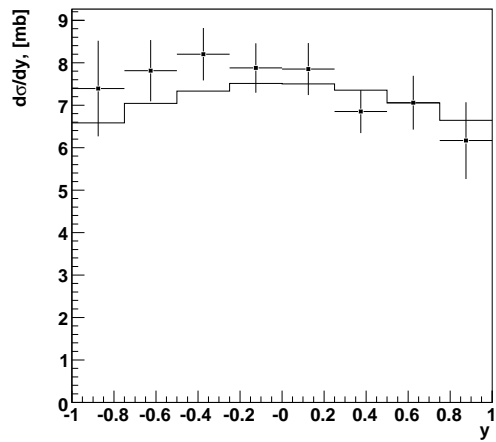}}
\end{picture}
\caption{\label{fig:crosssectrap}
Coherent $\rho^{0}$  production cross-section for the  minimum bias data
set  as a  function of  $y_{\rho^{0}}$  (black dots)  overlaid by  the
 $d\sigma/dy$  distribution predicted  by the KN  model~\cite{kn}
(solid line).}
\end{figure}

The  differential   cross-section  as  a  function   of  rapidity  for
$\rho^{0}$  photoproduction ($d\sigma/dy$)  obtained with  the minimum
bias sample is shown in Fig \ref{fig:crosssectrap}. 
The  distribution for each
rapidity bin was fit with Eq. \ref{eq:fitfunc} and the $\rho^{0}$ yield
extracted.  Also
shown   is  a   prediction  of   the  KN   model;
the other two models do not include nuclear excitation.

\begin{figure}[htb]
\begin{picture}(100,200)
\put(-60,0){\includegraphics{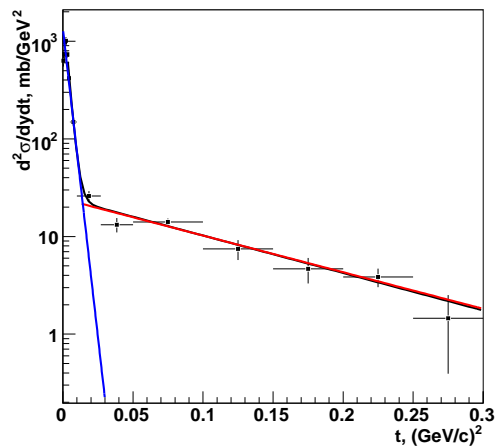}}
\end{picture}
\caption{\label{fig:crosssect} $\rho^{0}$  production cross-section as
  a function of the momentum transfer squared $t$, together  with the fit of Eq. \ref{eq:dsdt}.  The
  fit parameters are shown in Table~\ref{tab:fitincoherent}.}
\end{figure}

Fig. \ref{fig:crosssect} shows the  $\rho^{0}$ spectrum as a function of
$t$ for minimum  bias data.  The  efficiency correction
and luminosity normalization have been applied.

We do not observe the dip in the range $0.01 < t < 0.015$~(GeV/c)$^2$ predicted by FSZ~\cite{fsz200}.  
The $p_T$ of the $\rho^{0}$ meson  is the vector sum of the photon $p_T$
and the $p_T$ transferred by  the target nucleus.  Since the direction
of  the photon and  scattering $p_T$  are uncorrelated,  this addition
smears out the diffractive dips~\cite{KNpt}.

The     $d^{2}\sigma/dydt$     distribution     (averaged     over
$|y_{\rho^{0}}|<1$)  is  fit  to   a  sum  of  two  exponentials,  which
correspond to coherent and incoherent production:
\begin{equation}
{d^{2}\sigma\over {dydt}} = A_{\text{coh}}\exp{(-B_{\text{coh}}t)} + A_{\text{inc}}\exp{(-B_{\text{inc}}t)}.
\label{eq:dsdt}
\end{equation}

The simple fit function shown in Eq. \ref{eq:dsdt} has two drawbacks. 
The interference  between  $\rho^{0}$ photoproduction  on  the two  nuclei
reduces  $d^{2}\sigma/dydt$ at  small  $t$~\cite{KNpt,  nbk}, and,  in
fact, alters  the minimum bias  $t-$spectrum at the 20~$\%$  level for
$t<0.01$~(GeV/c)$^2$.

A fit over  all $t$ values yields exponentials (shown in the left column of Table I) 
that are  integrable to  give the total coherent cross-section 
(given are the parameters of the exponentials, not the cross-section).
 Because of the  interference at small  $t$, this fit  has a
poor $\chi^2/DOF$ of  $79.12/10$.  A second fit (shown in the right column of Table I) 
 is  performed over the $t$ range  from 0.002 to  0.3~(GeV/c)$^2$. It avoids the  region where
the interference is large and has a $\chi^2/DOF$ of $8.1/7$.
 This fit  yields a  nuclear slope with  accuracy comparable  to other
experiments.  Both  fits  give  similar  results  for  the  incoherent
production.

The incoherent slope, $B_{\text{inc}}$  = $8.8 \pm 1.0$ (GeV/c)$^{-2}$
has not previously been determined  in heavy ion collisions.  However, it
is comparable to the slope observed by STAR in $\text{dAu}$ collisions
~\cite{timosh}, and comparable  to the ZEUS result B$_{p}$ =  10.9 $\pm$ 0.3 (stat.)
$^{+1.0}_{-0.5}$ (syst.)~(GeV/c)$^{-2}$~\cite{breit}  and H1 result B$_{p}$
=   10.9   $\pm$   2.4   (stat.)  $\pm$   1.1   (syst.)~(GeV/c)$^{-2}$
~\cite{h1hera}  for  $\rho^{0}$  photoproduction  on  proton  targets  at
comparable $t$  values.  The HERA  data are at higher  $W_{\gamma N}$,
but the energy difference is not expected to introduce a large shift.

With  the  second  fit region we  find  the  coherent  production  slope  of
$B_{\text{coh}}$ =  $388~\pm~24$~(GeV/c)$^{-2}$,  obtained with the double
exponential  fit function.   For direct  comparison with  the previous
STAR result,  we also performed  a single exponential fit,  which gave
$B_{\text{coh}}$ =  $363~\pm~21$~(GeV/c)$^{-2}$ in  agreement with the
value  observed at 130~GeV,  $358~\pm~31$~(GeV/c)$^{-2}$~\cite{meis}.
These  numbers   are  not   directly  comparable  with   fixed-target
photoproduction data because  the photon flux in UPC photoproduction
is higher  on the side of  the target nearest the  photon emitter and
lower  on the  far  side of  the  target.  The  photon  flux falls  as
$1/r^2$, which leads to a slightly smaller apparent source size.

Despite these difficulties,
the  two   exponentials  in   Eq.  \ref{eq:dsdt}  were  integrated
analytically to find the total coherent and incoherent cross-sections.
This  approach neglects the  corrections  due to  the  loss of  incoherent
cross-section    when   the    coherent    cross-section   is    large
~\cite{bochmann}, but is useful for phenomenological comparisons.  For
$|y_{\rho^{0}}|<1$, we find the ratio
\begin{equation}\nonumber 
\sigma_{\text{incoherent}}/\sigma_{\text{coherent}}= 0.29 \pm 0.03~\text{(stat.)} \pm 0.08~\text{(syst.)}
\end{equation}
for events with mutual excitation ($XnXn$).

We  have  also  studied  the cross-sections  for  $\rho^{0}$  production
accompanied by single neutron  emission ($1n1n$) which is largely due
to  mutual excitation  of Giant  Dipole Resonances.   We did this by
fitting  the ZDC  spectra  in Fig.  \ref{fig:zdc}  and extracting  the
single   neutron    component from the fit. For   $|y_{\rho^{0}}|<1$, we find
$\sigma_{\text{incoherent}}^{1n1n}/\sigma_{\text{coherent}}^{1n1n}=
0.18$ $\pm$      $0.08$ (stat.)      $\pm$       $0.05$ (syst).       The       higher
$\sigma_{\text{incoherent}}/\sigma_{\text{coherent}}$  for  the $XnXn$
sample  may  signal  a  breakdown  of the  factorization  implicit  in
Eq.  \ref{eq:crossprob} because  the incoherent  $\rho^{0}$
production transfers  enough energy to disassociate  the target nucleus.
This effect would lead     to  additional   multiple     neutron     emission
~\cite{strikmanneutron}.

\begin{table}
\caption{\label{tab:fitincoherent}
Parameters for the fit to the $d^{2}\sigma/dydt$, Eq. \ref{eq:dsdt} }
\begin{ruledtabular}
\begin{tabular}{lll}
Parameter                    & $t$ range (0.,0.3)  &   $t$ range (0.002,0.3)   \\            
                             & (GeV/c)$^2$         & (GeV/c)$^2$ \\ \hline  
$A_{\text{coh}}$, mb/(GeV/c)$^{2}$  & 1050 $\pm$ 57  & 2307 $\pm$ 258 \\
$B_{\text{coh}}$, (GeV/c)$^{-2}$    & 256 $\pm$ 12  & 388 $\pm$ 24  \\
$A_{\text{inc}}$, mb/(GeV/c)$^{2}$  & 21.6 $\pm$ 2.4 &  24.8 $\pm$ 2.5 \\
$B_{\text{inc}}$, (GeV/c)$^{-2}$    & 7.9 $\pm$ 0.9 &  8.8 $\pm$ 1.0 \\

\end{tabular}
\end{ruledtabular}
\end{table}

\subsection{Total Cross-Sections}

\begin{figure}[htb]
\begin{picture}(100,200)
\put(-60,0){\includegraphics{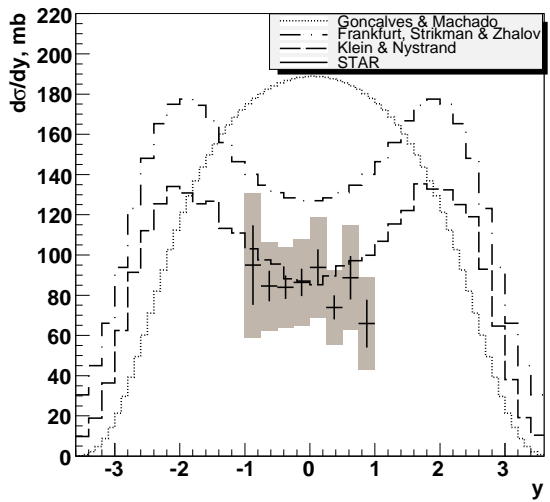}}
\end{picture}
\caption{\label{fig:finalcross} Comparison of theoretical predictions to the measured differential 
cross-section for coherent $\rho^{0}$ 
production. The statistical errors are shown by the solid vertical line at each data point. 
The sum of the statistical and systematic error bars is shown by the grey band.  
}
\end{figure}

We have compared three theoretical models to our measurements~\cite{ksjn,fsz,gm}. 
The comparison is shown in Fig. \ref{fig:finalcross}.

The    total    production    cross-section,
$d\sigma_{tot}/dy$,  is  obtained  by  scaling the  mutual  excitation
results with the scaling factors
$\sigma({\rho^{0}}_{0n0n})/\sigma({\rho^{0}}_{XnXn})$ and
$\sigma({\rho^{0}}_{0nXn})/\sigma({\rho^{0}}_{XnXn})$ as a function of
rapidity. 
 The scaling is needed because  the efficiency of the topology trigger,
which enters into the total cross-section, is only poorly known.
 Therefore the $\rho^{0}$ production  cross-section for the events with
mutual  excitation ($XnXn$) measured  with  the minimum  bias  sample was
extrapolated  based  on  the ratios  $\sigma(0n0n)/\sigma(XnXn)$=  7.1
$\pm$  0.3 (stat.) and  $\sigma(0nXn)/\sigma(XnXn)$ =3.5  $\pm$ 0.2 (stat.) which are
measured within the topology sample. 
Due  to the  limited  acceptance in  rapidity,  we cannot  distinguish
between the different theoretical  models based on the shape. However,
the  amplitude can  be used  to eliminate  models  which significantly
overestimate  the  total  production  cross-section  in  the  measured
rapidity range.

The cross-sections for coherent and incoherent production for 
$|y_{\rho^{0}}|<1$ accompanied by nuclear excitation are
$\sigma_{\text{coh}} (XnXn, |y_{\rho^{0}}|<1 )$ =  14.5   $\pm$ 0.7 (stat.)  $\pm$ 1.9 (syst.)~mb and 
$\sigma_{\text{inc}} (XnXn, |y_{\rho^{0}}|<1 )$ = 4.5 $\pm$ 0.5 (stat.) $\pm$ 0.6 (syst.)~mb.

Finding the total cross-sections requires an extrapolation to the
region $|y_{\rho^{0}}|>1$. This extrapolation is necessarily model dependent.  The KN
~\cite{kn} and FSZ~\cite{fsz200} calculations have similar
$d\sigma_{tot}/dy$ distributions and so a single extrapolation should
work well for them.  For the KN calculation, the extrapolation factor
from $\sigma(|y_{\rho^{0}}|<1)$ to $\sigma_{tot}$ is 2.2 $\pm$ 0.1 for the events
with nuclear break-up.  We assume that this factor is the same for
coherent and incoherent production. 
 The   coherent   production  cross-section
extrapolated to the full rapidity range is $\sigma_{\text{coh}} (XnXn,
\textrm{full-}y)$  =  31.9  $\pm$  1.5 (stat.) $\pm$  $4.5  \textrm{(syst.)}\  \textrm{mb}$.   The  total
cross-section is
\begin{equation}\nonumber 
\sigma_{\text{coh + inc}} (XnXn, \textrm{full-}y) = 41.4\ \pm\ 2.9\ \text{(stat.)}\ \pm\ 5.8\ \text{(syst.)}\ \textrm{mb}
\end{equation}
the $XnXn$ denotes multiple neutron emission due to nuclear dissociation.

For $\rho^{0}$ production accompanied by single neutron emission, we find, 
$\sigma_{\text{coh}}(1n1n, |y_{\rho^{0}}|<1 ) = 1.07\ \pm\ 0.08\ \text{(stat.)}\ \pm\ 0.14\ \text{(syst.)}\ \textrm{mb}$ and 
$\sigma_{\text{inc}}(1n1n, |y_{\rho^{0}}|<1 ) = 0.21\ \pm\ 0.09\ \text{(stat.)}\ \pm\ 0.03\ \text{(syst.)} \ \textrm{mb}$. 

The extrapolation factor from $|y_{\rho^{0}}|<1$ to $4\pi$ is assumed to be the same as
that for the $XnXn$ dataset (i.e. 2.2).
The total cross-section for single neutron emission is  
\begin{equation}\nonumber 
\sigma_{\text{coh + inc}} (1n1n, \textrm{full-}y) = 2.8\ \pm\ 0.3\ \text{(stat.)}\ \pm\ 0.4\ \text{(syst.)}\ \textrm{mb}
\end{equation}

Based on the ratio $\sigma({\rho^{0}}_{0n0n})/\sigma({\rho^{0}}_{XnXn})$, we find 
$\sigma_{\text{coh}}(0n0n,|y_{\rho^{0}}|<1) = 106\ \pm\ 5\ \text{(stat.)}\ \pm\ 14\ \text{(syst.)}\ \textrm{mb}$.

As  with  the  $XnXn$  data,  the extrapolation  to  $4\pi$  is  model
dependent.  For the  KN model, the extrapolation factor  is 3.7.  For
the FSZ model,  the factor would be 3.5, and  for the saturation model
GM~\cite{gm},  2.13.  The  KN and FSZ  model factors are  similar, what is remarkable since the predicted cross sections 
are differ by 60$\%$. Since the  KN $d\sigma/dy$ matches the  $XnXn$ data well, 
we adopt  an overall extrapolation factor of $3.7
\pm 0.1$.  With that, we find $\sigma_{\text{coh}}(0n0n, \textrm{full-}y)$ = 391
$\pm$ 18 (stat.) $\pm$ 55 (syst.) $\ \textrm{mb}$  and a total cross-section  of events
with $0n0n$ (coherent, incoherent) is
\begin{equation}\nonumber 
\sigma_{\text{coh+inc}}(0n0n, \textrm{full-}y) = 508\ \pm\ 24\ \text{(stat.)}\ \pm\ 71\ \text{(syst.)}\ \textrm{mb}
\end{equation}

It is also possible for a single nucleus to be excited, $0nXn$ in this
language. The $0nXn$ cross-section includes the possibility for either of the two nuclei to dissociate. 
 We have  checked that we get  symmetric results for  this channel when
the signals  are in the east or  west ZDC; those events  are added and
treated equally. 

This yields the total coherent cross-section 
$\sigma_{\text{coh}}(\text{AuAu} \rightarrow \text{Au}^{(*)}\text{Au}^{(*)}\rho^{0})$ = 530 $\pm$  19 (stat.) $\pm$ 57 (syst.)$\ \textrm{mb}$, 
and total cross-section (coherent, incoherent) 
\begin{eqnarray}\nonumber 
\sigma_{\text{coh+inc}}(\text{AuAu} \rightarrow \text{Au}^{(*)}\text{Au}^{(*)}\rho^{0}) &=& 
\nonumber \\
697\ \pm\ 25\ \text{(stat.)}\ \pm\ 73\ \text{(syst.)}\ \textrm{mb}. 
\nonumber
\end{eqnarray}

Table   \ref{tab:crossall}  summarizes   the  measured   coherent  and
incoherent production  cross-sections  and compares them  with results
obtained at $\sqrt{s_{NN}}$ = 130~GeV~\cite{meis}. 
  The measured 12\% increase
(with large errors) in the coherent photoproduction cross-section (going from 130 to 200 GeV
collisions) is much less than is predicted by all 3 models ~\cite{fsz}, ~\cite{kn} and ~\cite{gm}, which predict 
cross-section increases of between 70\% and 80\%. Our results
 at $\sqrt{s_{NN}}$ = 200 GeV/c support the $\rho^{0}$-nucleon cross-sections used in KN~\cite{kn}.

\begin{table*}
\caption{\label{tab:crossall}The total cross-section extrapolated to the full rapidity range for coherent $\rho^{0}$ production
  at $\sqrt{s_{NN}}$ =  200~GeV accompanied  by nuclear
  breakup and  without breakup compared with  previous measurements at
  $\sqrt{s_{NN}}$ = 130~GeV~\cite{meis}. The first error is statistical, the second is systematic.  }
\begin{ruledtabular}
\begin{tabular}{cccc}
Parameter                        & STAR at             & STAR at                                   & STAR at                                                    \\ 
                                 & $\sqrt{s_{NN}}$ = 130~GeV & $\sqrt{s_{NN}}$ = 200~GeV                  & $\sqrt{s_{NN}}$ = 200~GeV                                 \\
\hline
                                & coherent                     & coherent                   & coherent + incoherent          \\
$\sigma^{\rho^{0}}_{XnXn}$ (mb)   &  28.3 $\pm$ 2.0 $\pm$  6.3                       & 31.9 $\pm$ 1.5  $\pm$ 4.5  & 41.4 $\pm$ 2.9 $\pm$ 5.8   \\ 
$\sigma^{\rho^{0}}_{0nXn}$ (mb)     &  95   $\pm$ 60 $\pm$  25                    & 105  $\pm$ 5       $\pm$ 15   & 145 $\pm$ 7    $\pm$ 20        \\ 
$\sigma^{\rho^{0}}_{1n1n}$ (mb)     &  2.8  $\pm$ 0.5 $\pm$  0.7                    & 2.4  $\pm$ 0.2   $\pm$ 0.4  & 2.8 $\pm$ 0.3  $\pm$ 0.4     \\ 
$\sigma^{\rho^{0}}_{0n0n}$ (mb)      &  370  $\pm$ 170 $\pm$ 80                   & 391  $\pm$ 18      $\pm$ 55   & 508 $\pm$ 24   $\pm$ 71        \\ 
$\sigma^{\rho^{0}}_{total}$ (mb)     &  460  $\pm$ 220 $\pm$ 110                  & 530  $\pm$ 19      $\pm$ 57   & 697 $\pm$ 25   $\pm$ 73      \\ 
\end{tabular}
\end{ruledtabular}
\end{table*}

Several  sources of  systematic  error have  been  considered in  this
analysis. 
The main sources of the  systematic errors for the cross-section in the
rapidity range  $|y_{\rho^{0}}|<1$ are 10~$\%$ for the luminosity measurement, 3~$\%$ due to the different approaches for the background description 
 and 7~$\%$ for the applied cuts
and fit  function.  The major additional systematic  error for  the total
coherent  and  incoherent   production  cross-sections 
 is 6~$\%$ for the extrapolation  to   the  full  rapidity interval. The error is mainly due to the 
difference between extrapolation factor in KN and FSZ models. These
uncertainties were  added in quadrature to give  the systematic error
 for the total production cross-section.

\subsection{\label{sub5:sec3}$\rho^{0}$ Spin Density Matrix}

The  angular   distribution  of  the   $\pi^+$  and  $\pi^-$   in  the
$\gamma-$nucleon center  of mass  frame can be  used to  determine the
$\rho^{0}$  spin  density matrix  elements.   This  has previously  been
studied  in  $\gamma  p$  collisions  at  HERA~\cite{breit}.   There,
measurement of the  recoiling proton allowed the $\gamma  p$ center of
mass  frame to  be determined.   STAR does  not observe  the recoiling
proton, and so  we cannot separate the measured $p_T$ into  contributions from the
photon and  from the nucleon.   Furthermore, there is a  two-fold ambiguity
about  photon direction.  Because  of these  problems, we  perform our
analysis with  respect to the  $z$-axis (beam direction).  Since the laboratory  frame is
heavily boosted  with respect to the $\gamma  p$ center-of-mass frame,
this is  a good approximation; in the target frame, the $\rho^{0}$ direction is within
1-2 mrad of the beam axis.   $\Theta_{h}$ is defined as  the polar
angle between the  beam direction and the direction  of the $\pi^+$ in
the  $\rho^{0}$  rest  frame.   With  the two-fold  ambiguity  in  photon
 direction, the $+z$ and $-z$ directions are equivalent.  This does not
 affect two terms: $\cos^{2}(\Theta_h)$ and $\sin^{2}(\Theta_h)$, since they  
are symmetric around $\pi/2$ (i.e. around mid-rapidity). However the term   $\sin(2\Theta_h)$ 
is not symmetric around $\pi/2$ and therefore we are not sensitive to 
the interference between helicity states non-flip to single flip.  
 The azimuthal angle $\Phi_{h}$ is the angle between the decay
plane and the  $\rho^{0}$ production plane. The production  plane of the
$\rho^{0}$ contains the $\rho^{0}$ and a virtual photon.
 The    dependence   of   the    cross-section   on    $\Phi_{h}$   and
$\cos(\Theta_{h})$ can be written as follows~\cite{schil}:

\begin{eqnarray}\label{eq:fitfunchelicity}
 \frac{1}{\sigma}\frac{d^2\sigma}{d \cos(\Theta_{h}) d \Phi_{h}}&\!=\!&\frac{3}{4\pi} \cdot [\frac{1}{2}(1-r^{04}_{00})
\nonumber \\
&&+\frac{1}{2}(3r^{04}_{00}-1)\cos^{2}(\Theta_{h})
\nonumber \\
&&-\sqrt{2}\Re e[r^{04}_{10}]\sin(2\Theta_{h})\cos(\Phi_{h})
\nonumber \\
&&-r^{04}_{1-1}\sin^{2}(\Theta_{h})\cos(2\Phi_{h})].
\end{eqnarray}

The  three  independent  spin  density  matrix  elements  $r^{04}_{00},
r^{04}_{10},  r^{04}_{1-1}$  can  be  extracted  by  fitting  the  two
dimensional angular correlation. The  superscripts indicate contributions from
the   photon   polarization   states~\cite{zeus99}.    The   element
$r^{04}_{00}$  represents  the  probability  that  the  $\rho^{0}$  is
produced  with helicity 0  from a  photon with  helicity $\pm$  1. The
element  $r^{04}_{1-1}$ is  related to  the size  of  the interference
between the helicity non-flip and double flip and $\Re e[r^{04}_{10}]$
is related to the interference  of non-flip to single flip, where $\Re
e[r^{04}_{10}]$ stands for the real part of $r^{04}_{10}$. If helicity
conservation holds, then all three matrix elements will be close to
zero.

\begin{figure}[htb]
\begin{picture}(100,110)
\put(-80,-10){\includegraphics{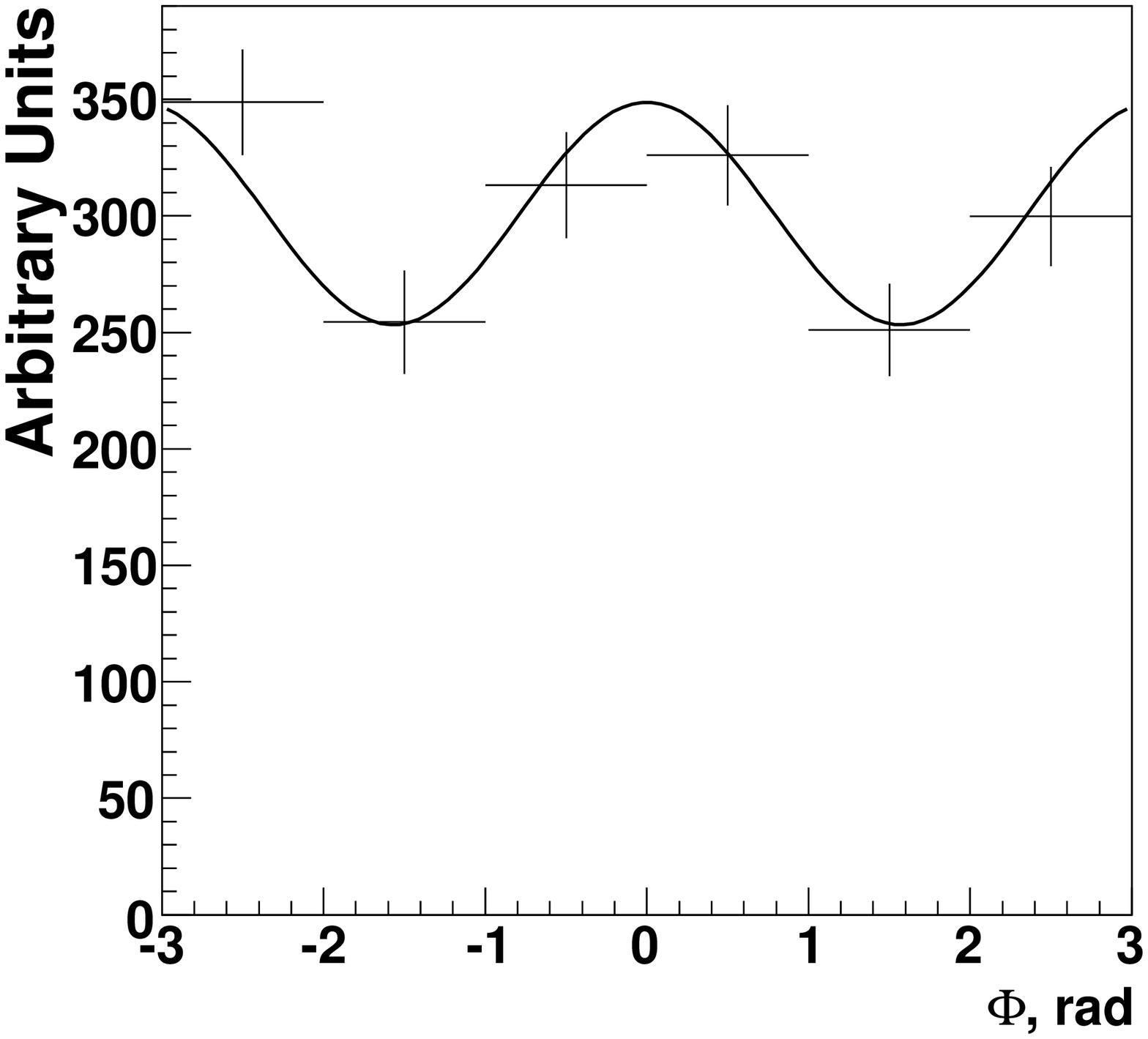}}
\put(50,-10){\includegraphics{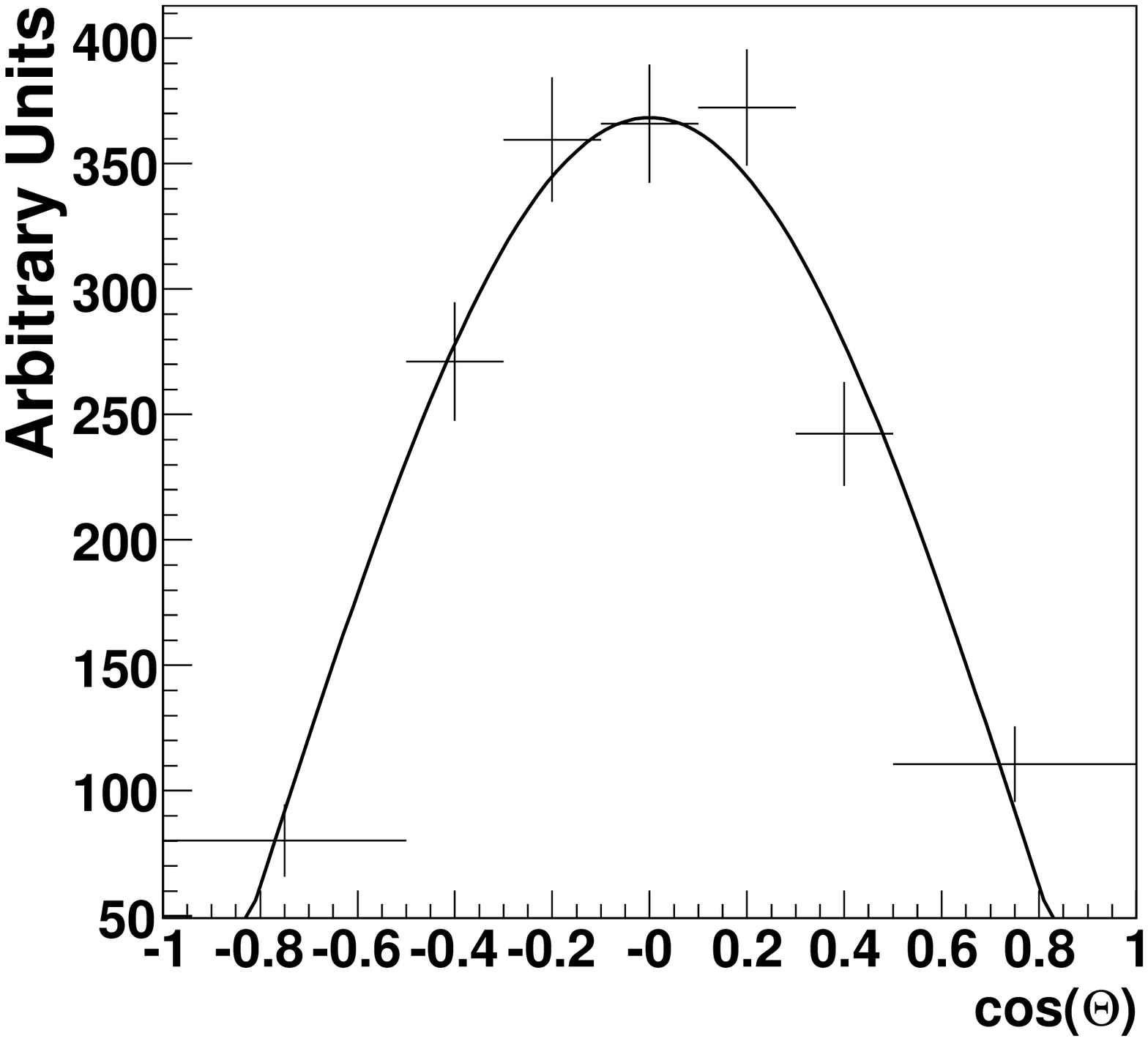}}
\end{picture}
\caption{\label{fig:phithetaraw}  Projections  of  the two  dimensional
  efficiency  corrected $\Phi_{h}$ vs  $cos(\Theta_{h})$ distributions
  obtained with the minimum bias data set. The solid line shows the result
  of     the    two-dimensional     fit    to     the     data    with
  Eq.   \ref{eq:fitfunchelicity}  and   the   coefficients  given   in
  Tab. \ref{tab:spinelem}.}
\end{figure}

Figure   ~\ref{fig:phithetaraw}  shows   the   $\Phi_{h}$   vs
$\cos(\Theta_{h})$  correlation fit for the  minimum  bias data
set.  The measured  spin density  matrix elements  are shown  in Table
\ref{tab:spinelem}.
 The method used is to fit  the invariant mass distributions in bins of
$\Phi_{h}$ and $\cos(\Theta_{h})$ to determine the yield in each bin.

The  background is  accounted for  in  the fitting  function with  non-scaled  
like-sign pairs  as described  in section  \ref{sub:sec2}. The
main  contribution  to  the  systematic  uncertainty  comes  from  the
background  subtraction.  It was  estimated  by  using an  alternative
approach where the scaled invariant mass distribution of the like-sign
pairs was subtracted from that of the opposite-sign pairs.
 An additional source  of systematic error  is the uncertainty due  to the
acceptance correction determined by a $\rho^{0}$ Monte Carlo simulation.
We estimate the systematic error by varying the bin size of $\Phi_{h}$
and $\cos(\Theta_{h})$.
The systematic error for the  spin density matrix elements is obtained
by adding the individual uncorrelated contributions in quadrature. The
measured $\rho^{0}$  helicity matrix elements  indicate that helicity
is  conserved within errors  as expected  based on  s-channel helicity
conservation.

\begin{table}
\caption{\label{tab:spinelem}Measured spin density matrix elements compared with ZEUS($\gamma p $) results. The first error is statistical, the second is systematic.}
\begin{ruledtabular}
\begin{tabular}{lcc}
Parameter & Fit result & $\gamma p $ experiment~\cite{breit} \\ \hline

$\chi^{2}/ndf$         &    26/21              &              \\
$r^{04}_{00}$      & -0.03 $\pm$ 0.03 $\pm$ 0.06  & 0.01 $\pm$ 0.01 $\pm$ 0.02    \\ 
$\Re e[r^{04}_{10}]$ & -  & 0.01 $\pm$ 0.01 $\pm$ 0.01  \\ 
$r^{04}_{1-1}$     & -0.01 $\pm$ 0.03 $\pm$ 0.05 & -0.01 $\pm$ 0.01  $\pm$ 0.01\\

\end{tabular}
\end{ruledtabular}
\end{table}

\section{\label{sec4}Conclusion}

Photoproduction  of $\rho^{0}$ mesons  has been  measured in  the STAR
detector   at   RHIC  in Au-Au collisions   at
$\sqrt{s_{NN}}$   =  200~GeV.   Coherent  and   incoherent  $\rho^{0}$
photoproduction  has  been  observed and photoproduction  of  the
$\rho^{0}$ mesons  is observed  with and without  accompanying Coulomb
nuclear   excitations.  The   measured  increase   with energy in the total cross-section for photoproduction is
much slower than proposed in~\cite{fsz200} and~\cite{gm}. 
 However, the Klein and Nystrand model~\cite{ksjn}  is able to describe the data for
two energy points $\sqrt{s_{NN}}$ = 130 and 200~GeV.

The differential cross-section for photoproduction has been studied as
a   function  of  $t$,   $y_{\rho^{0}}$  and   $M_{\pi^{+}\pi^{-}}$.   The
$d^{2}\sigma/dydt$  distribution  was fit  with  a double  exponential
function to isolate the incoherent part of the $\rho^{0}$ production cross-section.

The ratio of $\pi^{+}\pi^{-}$  to direct $\rho^{0}$ production ($|B/A|$) has
been   studied   with   respect   to  polar angle,   azimuthal   angle   and
$y_{\rho^{0}}$;   no  dependence  has   been  observed   as  predicted in reference
~\cite{ryskin}.

The  $r^{04}_{00}$  and  $r^{04}_{1-1}$  spin
density matrix elements for the $\rho^{0}$ meson were measured.
The   small  values   of  $r^{04}_{00}$  and
$r^{04}_{1-1}$ indicate  that helicity  is conserved within  errors as
expected based on s-channel helicity conservation (SCHC), and therefore
we see no evidence for $\rho^{0}$ photoproduction involving spin flip.

\begin{acknowledgments}
We  thank the  RHIC Operations  Group and  RCF at  BNL, and  the NERSC
Center at LBNL  for their support. This work was  supported in part by
the Offices of  NP, HEP and EPSCoR within the U.S. DOE  Office of Science; the
U.S. NSF; the BMBF of Germany; CNRS/IN2P3, RA, RPL, and EMN of France;
EPSRC of the United Kingdom; FAPESP of Brazil; the Russian Ministry of
Sci. and Tech.; the Ministry of  Education and the NNSFC of China; IRP
and GA  of the Czech Republic,  FOM of the Netherlands,  DAE, DST, and
CSIR of the Government of India; Swiss NSF; the Polish State Committee
for Scientific  Research; Slovak Research and  Development Agency, and
the Korea Sci. $\&$ Eng. Foundation.
\end{acknowledgments}

\end{document}